\documentclass[12pt]{article}
\pdfoutput=1

\usepackage[]{hyperref}
\usepackage[]{putex}
\usepackage[utf8]{inputenc}
	
\usepackage[table]{xcolor}

\usepackage{amsmath,amsthm,amssymb}
\usepackage{tikz}
\usepackage{tikz-cd}
\usetikzlibrary{decorations.markings}
\usetikzlibrary{calc}
\usepackage{subcaption}
\usepackage{enumerate}
\usepackage{enumitem}
\usepackage{changepage}
\usepackage{multirow}
\usepackage{bm}
	
\definecolor{lightgray}{gray}{0.9}
\definecolor{lightgray2}{gray}{0.95}
\definecolor{colour1}{rgb}{0, 0, 8/10}
\definecolor{colour2}{rgb}{0.4, 0.5, 1}
\definecolor{colour3}{rgb}{0.7, 0.85, 1}
\definecolor{colour4}{rgb}{0.922526, 0.385626, 0.209179}

\definecolor{colour5}{rgb}{0.560181, 0.9, 0.194885}
\definecolor{colour6}{rgb}{0.260181, 0.691569, 0.194885}
\definecolor{colour7}{rgb}{0.847624, 0.37816, 0.614037}
\definecolor{colour8}{rgb}{0.647624, 0.37816, 0.614037}

\title{Cutting the Coon Amplitude
}

\authors{Christian Baadsgaard Jepsen}
\institution{SCGP}{Simons Center for Geometry and Physics, Stony Brook University, Stony Brook, NY, 11794}

\abstract{
The Coon amplitude is a $q$-deformed generalization of the Veneziano amplitude exhibiting a semi-infinite sequence of poles that converge on an accumulation point, from which a branch cut emerges. A number of recent papers have provided compelling evidence that the residues of this amplitude satisfy the positivity requirements imposed by unitarity. This paper investigates whether positivity is also satisfied along the branch cut. It is found that positivity violations occur in a region of the branch cut exponentially close to the accumulation point according to a scale set by $q$. The closing section of the paper discusses possible interpretations of this fact and strategies for excising negativity from the partial wave coefficients. 

An appendix presents derivations of instrumental identities relating the $q$-gamma and $q$-polygamma functions to the Weierstrass elliptic and quasiperiodic functions.
}

\begin{document}

\maketitle
\tableofcontents

\section{Introduction and Summary}

\begin{figure}[h]
    \centering
\begin{align*}
\begin{matrix}\text{
		\includegraphics[scale=0.8]{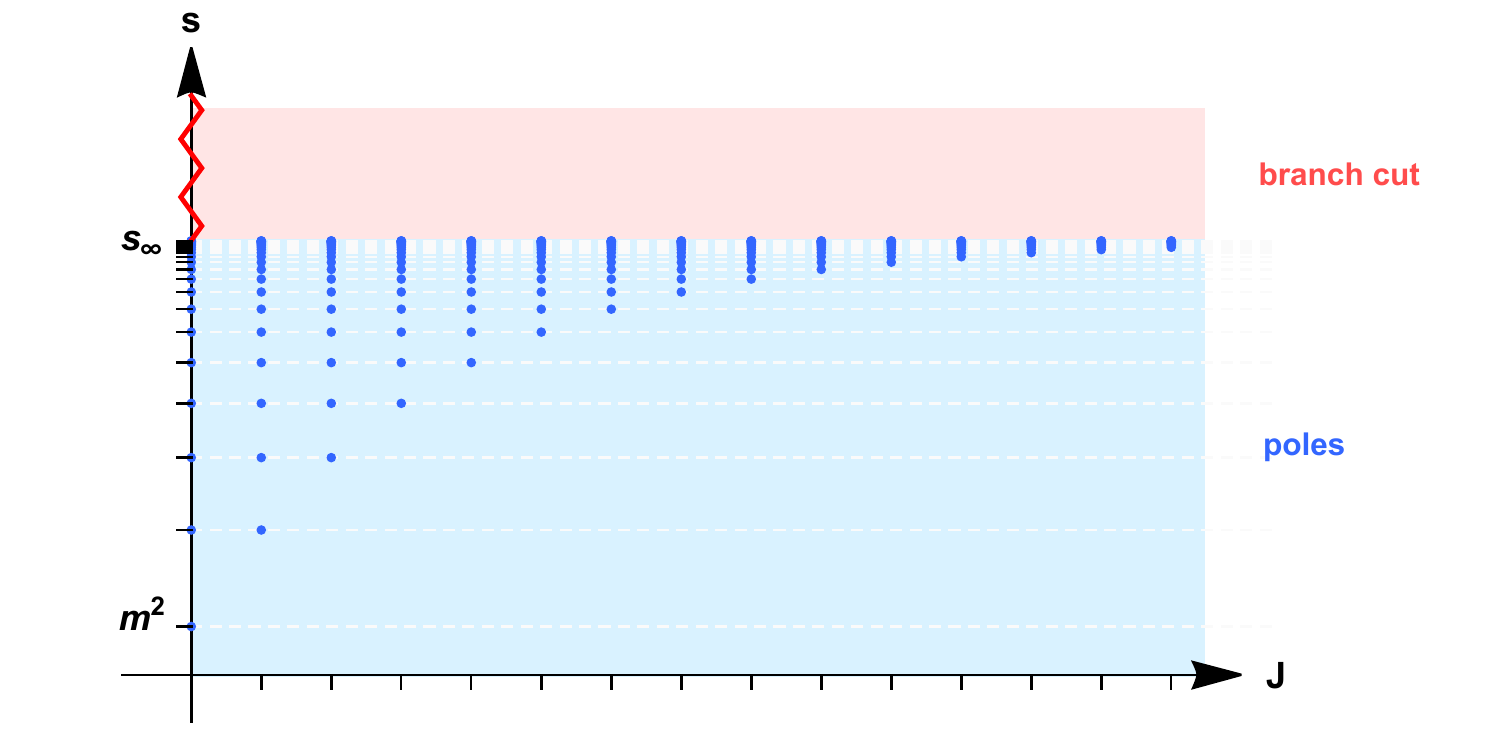}
}\end{matrix}
\end{align*}
    \caption{Regge trajectories of the Coon amplitude.}
    \label{fig:Regge}
\end{figure}

It is a curious fact of theoretical physics that the discovery of scattering amplitudes sometimes precede the knowledge of the physical processes they describe. The most important example is Gabriele Veneziano's 1968 discovery in Ref.~\cite{veneziano1968construction} of the amplitude that bears his name, and which gave birth to string theory. In the year following Veneziano's discovery, Darryl D. Coon in Ref.~\cite{coon1969uniqueness} put forward a $q$-deformation of the Veneziano amplitude: 
\begin{align}
\label{Afirst}
\mathcal{A}_q(s,t)\sim\,&
\prod_{n=0}^\infty
\frac{(\sigma\tau-q^n)}{(\sigma-q^n)(\tau-q^n)}\,,
\end{align}
where $q$ is a real number between zero and one, and $\sigma$ and $\tau$ are the following linear functions of Mandelstam invariants $s$ and $t$: 
\begin{align}
\label{eq:sigmaAndTau}
\sigma=1-(s-m^2)(1-q)\,, 
\hspace{15mm}
\tau=1-(t-m^2)(1-q)\,.
\end{align}
A peculiar characteristic of Coon's amplitude is that its spectrum exhibits a semi-infinite sequence of poles at values $s_n$ which are situated according to the $q$-integers $[n]_q$,
\begin{align}
\label{eq:poles}
s_n=m^2+[n]_q=m^2+\frac{1-q^n}{1-q}\,, \hspace{10mm} n\in \mathbb{N}_0\,,
\end{align}
and which converge on an accumulation point $s_\infty$ at finite $s$, 
\begin{align}
s_\infty = m^2+\frac{1}{1-q}\,.
\end{align}
Contrary to initial beliefs, the residues of Coon's original amplitude are not polynomial,\footnote{Presumably, Ref.~\cite{coon1969uniqueness} claimed that the residues of $\mathcal{A}_q(s,t)$ are polynomial because that appears to be the case when telescoping the infinite product at the residue. However, telescoping is only valid after dividing numerator and denominator in \eqref{Afirst} with $\sigma\tau$, which makes them separately convergent.} which means that the sum over partial waves on the pole does not terminate; in other words, in a putative theory associated to the amplitude, the spectrum would include an infinite tower of higher-spin particles for each of the infinite poles. As observed by Coon, Sukhatme, and Tr\^an Thanh V\^an in 1973 in Ref.~\cite{coon1973duality}, this situation can be remedied by writing down a modified amplitude:\footnote{Ref.~\cite{coon1973duality} attributed the modified Coon amplitude to an upcoming paper by Darryl Coon and Marshall Baker. It appears this paper was never published.}
\begin{align}
\label{eq:Acoon}
A_q(s,t)=\,&(q-1)\exp\Big(\frac{\log \sigma\log\tau}{\log q}\Big)
\prod_{n=0}^\infty
\frac{(\sigma\tau-q^n)(1-q^{n+1})}{(\sigma-q^n)(\tau-q^n)}\,.
\end{align}
The term ``Coon amplitude" is usually, including in this paper, applied to $A_q(s,t)$ rather than $\mathcal{A}_q(s,t)$.
The exponential factor in \eqref{eq:Acoon} has been chosen precisely such as to render all residues polynomial. Additional $s$- and $t$-independent factors are mostly conventional but result in the following limits:
\begin{align}
\lim_{q\rightarrow 1}A_q(s,t)=\,&-\frac{\Gamma(m^2-s)\Gamma(m^2-t)}{\Gamma(2m^2-s-t)}\,,
\\
\lim_{q\rightarrow 0}A_q(s,t) 
=\,&
\frac{1}{s-m^2}+\frac{1}{t-m^2}-1\,.
\end{align}
The Regge trajectories of the Coon amplitude are depicted Figure \ref{fig:Regge}. Because the spectrum is determined by $q$-integers rather than integers, the Regge trajectories are not linear but grow with a rate that decreases exponentially so that all trajectories converge on the accumulation point. 

Despite retaining crossing-symmetry and the benign high-energy behaviour that characterizes string theory amplitudes and despite indications of unitarity in Refs.~\cite{coon1974dual,fairlie1995fresh}, the Coon amplitude has eluded a physical interpretation up to the present time. Recent years, however, and last year especially, have witnessed a surge of interest in the Coon amplitude along with a number of papers that have systematically explored questions of unitarity. Ref.~\cite{caron2017strings} demonstrated that for any unitary theory of scalars weakly coupled to particles of spin greater than two, the scalar 4-point scattering amplitude must necessarily tend to the Veneziano amplitude in the (unphysical) limit of large $s$ and $t$, assuming crossing symmetry, polynomial boundedness, asymptotic Regge behaviour, and meromorphicity. The authors of Ref.~\cite{caron2017strings} pointed out the Coon amplitude as a notable example of an amplitude whose residues are sums of Legendre polynomials with positive coefficients, as stipulated by unitarity, but which does not meet the requirement of meromorphicity in consequence of its accumulation point and branch cut. Ref.~\cite{figueroa2022unitarity} (whose choice of normalization has been adopted in equation \eqref{eq:Acoon}) computed the Wilson coefficients associated with the Coon amplitude and refined its partial wave analysis by providing an analytic proof of regions in the space of $q$, $m$ where all the partial wave coefficients are positive for any $d$ and by presenting numerical evidence of critical dimensions for other values of $q$ and $m$.\footnote{See Ref.~\cite{chakravarty2022positivity} for a calculation of the partial wave coefficients of the leading Regge trajectory.} Further evidence for partial wave unitarity on the poles was furnished by Ref.~\cite{bhardwaj2022unitarity}, which employed a combination of $q$-analysis and recent techniques for studying the unitarity of string theory amplitudes developed in Ref.~\cite{arkani2022unitarity}. The application of $q$-analysis to the Coon amplitude had also been a feature of  Ref.~\cite{geiser2022properties}, which elaborated on the low-energy analysis of the Coon amplitude performed in Ref.~\cite{figueroa2022unitarity} and presented the Coon amplitude in the equivalent form 
\begin{align}
\label{eq:AcoonAgain}
A_q(s,t)=-q^{\frac{\log \sigma\log\tau}{\log^2 q}} \frac{\Gamma_q(-\frac{\log\sigma}{\log q})\Gamma_q(-\frac{\log\tau}{\log q})}{\Gamma_q(-\frac{\log\sigma+\log\tau}{\log q})}\,,
\end{align}
where $\Gamma_q(z)$ is the $q$-deformed gamma function, defined for complex $q$ with $|q|<1$ as
\begin{align}
\Gamma_q(z) \equiv (1-q)^{1-z} \prod_{n=0}^\infty \frac{1-q^{n+1}}{1-q^{n+z}}\,.
\end{align}
The possibility that the Coon amplitude may furnish an example of a legitimate unitary scattering amplitude with an accumulation point spectrum harmonizes with a conjecture advanced in Ref.~\cite{chiang2022non} that accumulation points are a quality of generic gravitational effective field theories and raises the question which further such amplitudes may yet be found.
An additional example of a $q$-family of amplitude was put forward already in Ref.~\cite{coon1973duality}, which $q$-deformed the amplitude of Lovelace \cite{lovelace1968novel} and Shapiro \cite{shapiro1969narrow}, but Ref.~\cite{coon1974dual} soon after pointed out that the $J=0$ states of the deformed amplitude contain ghosts, although Ref.~\cite{fernandez2022cornering} much later on reported numerical indications that the $J>0$ states are ghost-free. The potential existence of unitary continuous families of scattering amplitudes carries important implications for the $S$-matrix bootstrap and the associated exclusion plots, like those of Refs.~\cite{albert2022bootstrapping,chen2022nonperturbative,fernandez2022cornering}, especially in light of Ref.~\cite{cheung2023stringy}, which adopted an analytic bootstrap approach and presented a bottom-up derivation of the Coon amplitude, after inputting assumptions of crossing symmetry, vanishing Regge limit, finite spin-exchange, and a $q$-integer spectrum.  

Addressing the question of whether the Veneziano amplitude admits yet further generalizations, Refs.~\cite{cheung2023veneziano} and \cite{geiser2022generalized} undertook methodical searches for new amplitudes and formulated discrete generalizations of the Coon amplitude also exhibiting accumulation point spectra, polynomial residues, polynomially bounded high-energy behaviours, and poles with positive partial wave coefficients. Meanwhile, as observed in Refs.~\cite{geiser2022properties,cheung2023veneziano,geiser2022generalized}, attempts to generalize and $q$-deform closed string amplitudes are hurdled substantial hindrances, a fact suggesting that string endpoints may play a crucial role in a potential string theory interpretation of the Coon amplitude. Further support for this supposition was provided by Ref.~\cite{maldacena2022accumulation}, which demonstrated that the scattering amplitude for open strings ending on D-branes in AdS also has an accumulation point in its spectrum, although the spectrum differs from that of the Coon amplitude at large spin for energies close to the accumulation point. But a concrete physical theory that can reproduce the Coon amplitude has yet to be found. Furthermore, while a number of recent papers have carried out numerous non-trivial checks of the unitarity Coon amplitude, these checks do not exhaust the requirements that must be met in order for $A_q(s,t)$ to be unitary. In particular, the modification of Coon's original amplitude that rendered the residues polynomial came at a price. For the exponential factor that was introduced into \eqref{eq:Acoon} gives rise to a branch cut emanating from the accumulation point, and unitarity requires this cut to satisfy the generalized optical theorem. The discrete set of conditions that have been checked by performing a partial wave analysis on the poles must be supplemented by a continuous set of conditions everywhere along the cut. The present paper undertakes to investigate whether these additional conditions are met. A summary of the remainder of the paper is as follows:
\begin{itemize}
    \item[---]  Section \ref{sec:Cut} studies the imaginary part of the Coon amplitude,
\begin{align}
\label{eq:imPart}
\text{Im}\,A_q(s,t) = \lim_{\epsilon \rightarrow 0^+}\frac{1}{i}\Big(A_q(s+i\epsilon,t)-A_q(s-i\epsilon,t)\Big)\,,
\end{align}
    and analyzes the constraints imposed on this quantity by unitarity. First we briefly discuss the imaginary part due to the poles of the amplitude and review the exclusion regions determined from partial wave analysis by decomposing the residues into sums over Gegenbauer polynomials. Subsequently we turn to the imaginary part of the amplitude due to its branch cut and study it from two perspectives: by analyzing the branch cut discontinuity and its $t$-derivatives in the forward limit $t\rightarrow 0$, and by numerically evaluating asymptotic partial wave coefficients in the limit as $s$ approaches $s_\infty$ from above. By the first method, we find that the Coon amplitude does not decompose into a positively weighted sum over Gegenbauer polynomials for any $m^2$ when $q$ is less than 0.52, a value that can only increase with improved computing power; and by the second method we find that positivity is violated in every case we check, specifically for $m^2\in \{-1,0,1/3\}$ and $q\in\{0.1,0.2,0.3,0.4,0.5,0.6,0.7,0.8,0.9\}$.
    \item[---]  Section \ref{sec:Cure} discusses potential strategies for and challenges to    
    further modifying the Coon amplitude so as to repair the positivity violations on the branch cut without spoiling the desirable pole structure.
    \item[---]  Appendix \ref{sec:App} derives a number of identities that relate the $q$-gamma function $\Gamma_q(z)$ and the $q$-polygamma function $\psi_q^{(n)}(z)$ to the Weierstrass functions $\wp(z|\omega_1,\omega_2)$, $\zeta(z|\omega_1,\omega_2)$, and $\sigma(z|\omega_1,\omega_2)$ in the special case when $\omega_1$ is real and $\omega_2$ is imaginary, including a $q$-generalization of Euler's reflection formula for the gamma function,
    \begin{align}
    \label{eq:reflection}
\Gamma(1-z)\Gamma(z) = \frac{\pi}{\sin(\pi z)}\,.
    \end{align}
    A consequence of these identities is a pair of formulas crucial to the branch cut analysis of section \ref{sec:Cut}. 
   
\end{itemize}

\section{Cutting the Coon Amplitude}
\label{sec:Cut}

Consider the scattering of scalar particles: $A(p_1)+B(p_2)\rightarrow A(p_3)+B(p_4)$. Unitarity of the $S$-matrix for this scattering process simply amounts to the equation $S^\dagger S = 1$. Famously, this condition entails that the $T$-matrix defined via $S=1+iT$ satisfies the relation
\begin{align}
\label{eq:T}
i(T^\dagger - T) = T^\dagger T\,. 
\end{align}
This matrix equation can be diagonalized by going to a partial wave basis for $T(s,t)$, described in general $d$ by Gegenbauer polynomials $ P_J^{(d)}(\cos\theta)=C_J^{(\frac{d-3}{2})}(\cos\theta)$ as functions of the scattering angle $\theta$. In the diagonal basis, the right-hand side of \eqref{eq:T} is an absolute square, which implies a positivity condition on the imaginary part of $T$. In general, unitary imposes a more powerful constraint than positivity by dictating a precise relation between the imaginary parts and absolute squares of the partial wave coefficeints of $T$, see for example \cite{correia2021analytical} for a detailed review, but for amplitudes that are only known to leading order in perturbation theory, we must content ourselves with positivity.

While unitarity of a given theory in principle only imposes positivity constraints on full amplitudes, for all known physical open string amplitudes, rather than relying on a cancellation in the full amplitude, unitarity is present already at the level of partial amplitudes,\footnote{ For example, the residues at even non-negative values of $\alpha' s$ in the partial Veneziano amplitude do carry purely positive Gegenbauer coefficients, despite the fact that these residues cancel in the full amplitude.} thereby allowing for the presence of Chan-Paton factors associated to string endpoints. Similarly, studies of the unitarity of the Coon amplitude focus on the presence and absence of unitarity for the partial amplitude $A_q(s,t)$ rather than for a prospective full amplitude like the sum $A_q(s,t)+A_q(s,u)+A_q(t,u)$.

For real values of $s$ and $t$, a scattering amplitude only becomes imaginary on a pole or on a branch cut. Both of these phenomena occur in the case of the Coon amplitude. The imaginary part on the poles is determined from the distributional fact that 
\begin{align}
\lim_{\epsilon\rightarrow 0^+}\frac{1}{i}\Big(\frac{1}{s-s_n+i\epsilon}-\frac{1}{s-s_n-i\epsilon}\Big)
=-2\pi\delta(s-s_n)
\,,
\end{align}
while the imaginary part on a branch cut is determined by the discontinuity $\text{Disc}A_q(s,t)$ on winding once around the cut. Picking the direction of winding to be counter-clockwise, we take $\text{Disc}A_q(s,t)$ to equal the value of $A_q(s,t)$ below the branch cut minus its value above. Assuming $t$ is not valued on a pole or on the branch cut, the imaginary part of the Coon amplitude, as given in \eqref{eq:imPart}, then equals the following:
\begin{align}
\text{Im}\,A_q(s,t)
=-2\pi\sum_{n=0}^\infty \delta(s-s_n)\, \underset{\,s=s_n}{\text{Res}}\,A_q(s,t)
-\theta(s-s_\infty)\,\frac{1}{i}\,\text{Disc}A_q(s,t)
\,,
\end{align}
where the $s$-channel residues and the branch cut discontinuity of the Coon amplitude are given by
\begin{align}
\label{eq:res1}
&\underset{s=s_n}{\text{Res}}\,A_q(s,t)\hspace{3.6mm}=
q^n
\prod_{l=1}^n
\frac{\tau-q^{-l}}{1-q^{-l}}\,,
\\
\label{eq:res2}
&\frac{1}{i}\,\text{Disc}A_q(s,t)
=2(q-1)|\sigma|^{\frac{\log\tau}{\log q}}\sin\Big(\pi\frac{\log\tau}{\log q}\Big)
\prod_{n=0}^\infty
\frac{(1-\frac{q^n}{\sigma\tau})(1-q^{n+1})}{(1-\frac{q^n}{\sigma})(1-\frac{q^n}{\tau})}\,.
\end{align}
with the locations $s_n$ of the poles given in equation \eqref{eq:poles}. Once we expand \eqref{eq:res1} and \eqref{eq:res2} in a partial wave basis,
\begin{align}
&\underset{s=s_n}{\text{Res}}\,A_q\Big(s,\frac{(s-4m^2)(\cos\theta-1)}{2}\Big)
\hspace{4mm}=\sum_{J=0}^{n} a_J^{(d)}(s_n)\, P_J^{(d)}(\cos\theta)\,,
\\
\label{eq:discDecomp}
&\frac{1}{i}\,\text{Disc}\,A_q\Big(s,\frac{(s-4m^2)(\cos\theta-1)}{2}\Big)
=\sum_{J=0}^{\infty} a_J^{(d)}(s)\, P_J^{(d)}(\cos\theta)\,,
\end{align}
unitarity stipulates, for our choice of overall normalization of $A_q(s,t)$, that all partial wave coefficients be non-negative:
\begin{align}
\label{eq:coeffIneq1}
&\forall n\in\mathbb{N}_0:\,\, \forall J\leq n:\,\,\,\, a_J^{(d)}(s_n)\geq 0 \,,
\\[4pt]
\label{eq:coeffIneq2}
&\forall s>s_\infty:\,\, \forall J\in\mathbb{N}_0:\,\, a_J^{(d)}(s)\geq 0 \,.
\end{align}
Branch cuts and the condition \eqref{eq:coeffIneq2} are abnormal in the context of tree-level string theory, but for all known stringy tree-amplitudes, versions of the criterion \eqref{eq:coeffIneq1} are satisfied throughout their semi-infinite sequences of poles when the target space dimensionality is at or below the critical dimension.\footnote{See Ref.~\cite{arkani2022unitarity} for recent progress towards a direct proof of this fact, and see also Ref.~\cite{caron2017strings} for an alternative derivation of the positivity criterion for tree-amplitude residues.} Typically, the strongest unitarity bounds are obtained from the lowest-lying non-constant residues. For the Coon amplitude, the first non-constant residue is the residue of the pole at $s=s_1$. The two partial wave coefficients associated to this residue are given by
\begin{align}
a_0^{(d)}(s_1)=q\frac{2-(1-m^2)q}{2}\,,\hspace{20mm}
a_1^{(d)}(s_1)=\frac{1-3m^2}{2(d-3)}q^2\,.
\end{align}
These coefficients respectively impose the rigorous unitarity constraints $1-\frac{2}{q} \leq m^2$ and $m^2\leq \frac{1}{3}$ for any value of $d$ greater than 3. In the range $-1 \leq m^2 \leq \frac{1}{3}$, Ref.~\cite{figueroa2022unitarity} was able to formulate a proof that all coefficients $a_J^{(d)}(n)$ are non-negative for any $d>3$ provided that $q<q_\infty(m^2)$, where
\begin{align}
q_\infty(m^2)=\frac{m^2-3+\sqrt{9+2m^2+m^4}}{2m^2}\,,
\end{align}
whereas for $q>q_\infty(m^2)$ Ref.~\cite{figueroa2022unitarity} reports numeric evidence of critical dimensions above which unitarity is violated and below which all the positivity conditions of the poles are met.

Unlike the positivity conditions on the poles of the Coon amplitude, the positivity conditions \eqref{eq:coeffIneq2} on the branch cut have remained largely unchecked.\footnote{In the case of the branch cut of the $q$-deformed Lovelace-Shapiro amplitude, Ref.~\cite{fernandez2022cornering} reports that numerical tests show that positivity is violated for $J = 0$ but suggest that it is satisfied for $J > 0$.
}
In the remainder of this section, we will perform such checks in two ways. The coefficients $a_J^{(d)}(s)$ can be extracted through the use of the orthogonality relation for the Gegenbauer polynomials and then checked for positivity. The integration involved in this procedure is complicated but can be carried out numerically case by case for different values of $q$, $m^2$, and $d$. We adopt this approach in subsection \ref{subsec:Numeric}. A study of the forward limit of the branch cut discontinuity and its derivatives provides a method of checking unitarity in any number of dimensions in one stroke and will be the subject of the next subsection. 

\subsection{Positivitity in the forward limit?}

The Gegenbauer polynomials have the property 
 that
\begin{align}
\frac{d^N}{dx^N}P_J^{(d)}(x)\Big|_{x=1}\geq 0\,,
\end{align}
for all $N \in \mathbb{N}_0$ and $d\geq 3$. Therefore, if all the coefficients $a_J^{(d)}(s)$ in the partial wave decomposition \eqref{eq:discDecomp} are non-negative, we find, by differentiating 
with respect to $\cos\theta$ at fixed $s$ and applying the chain-rule, that\footnote{In deriving the inequality \eqref{eq:ineq1}, we are commuting the order of operations in performing differentiation and summation. The identification $\frac{d}{dx}\sum_n f_n(x)=\sum_nf'(x)$ is only guaranteed to be valid if the sum $\sum_n |f_n'(x)|$ converges. In our case, the coefficients $a_J^{(d)}(s)$ decay exponentially at large $J$, while the maximal value of any derivative of $P_J(\cos\theta)$ grows polynomially, so this subtlety does not pose an issue. More precisely, the maximal value of the $N$-fold derivative of the Gegenbauer polynomial $C_n^{(\alpha)}(x)$ over the interval $x\in (-1,1)$ grows as $x^{2N+2\alpha-1}$.}
\begin{align}
\label{eq:ineq1}
(s-4m^2)^N\,\frac{\partial^N}{\partial t^N}\text{Disc}\big[\text{Im}\,A_q(s,t)\big]\Big|_{t=0}\geq 0\,.
\end{align}
On the branch cut, we have that 
\begin{align}
s-4m^2>s_\infty-4m^2=\frac{1}{1-q}-3m^2\,. 
\end{align}
From the analysis of the poles of the Coon amplitude, we know that we must require $m^2\leq \frac{1}{3}$ for unitarity, which implies that $s-4m>0$ on the branch cut. The inequality \eqref{eq:ineq1} then simplifies to\footnote{For ways to extend the inequality \eqref{eq:ineq2} beyond the forward limit in the context of general scalar theories, we refer the reader to \cite{de2017positivity}.}
\begin{align}
\label{eq:ineq2}
\frac{\partial^N}{\partial t^N}\text{Disc}\big[\text{Im}\,A_q(s,t)\big]\Big|_{t=0}\geq 0\,.
\end{align}
Below we will check this inequality for the first several values of $N$. 

\subsection*{$\bullet\,\,N = 0$}

\begin{figure}[h]
    \centering
\begin{align*}
\begin{matrix}\text{
		\includegraphics[width=0.4\textwidth]{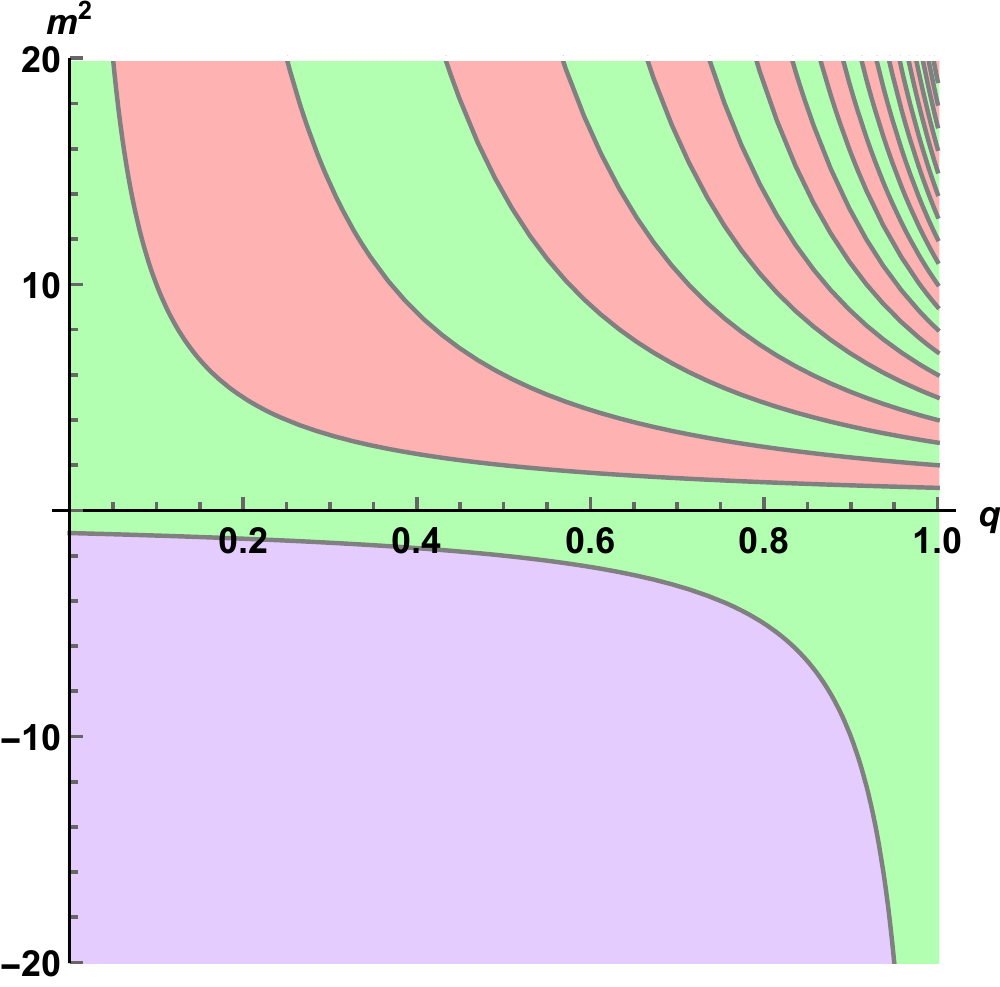}
}\end{matrix}
\hspace{14mm}
\begin{matrix}\text{
		\includegraphics[width=0.4\textwidth]{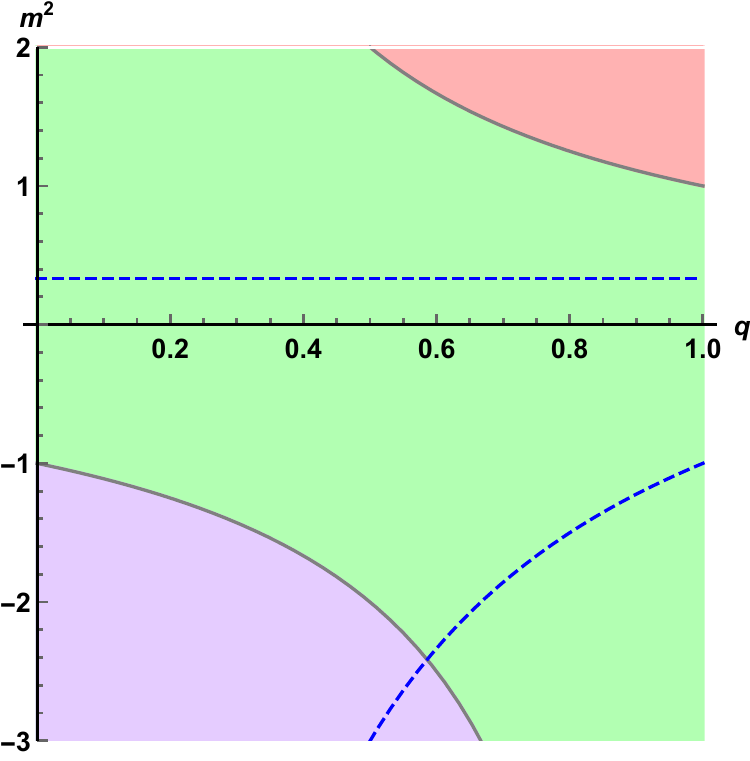}
}\end{matrix}
\end{align*}
    \caption{Green: Regions of parameter space where the branch cut discontinuity is positive everywhere on the branch cut. Red: Regions for which positivity is violated. Purple: Region where the branch cut is complex. The blue dotted lines in the zoomed-in plot on the right indicate the bounds $1-\frac{2}{q}\leq m^2$ and $m^2<\frac{1}{3}$ coming from the residues.}
    \label{fig:Disc}
\end{figure}

Before asking if the imaginary branch cut discontinuity is positive or negative, we should ask if the discontinuity is even purely imaginary in the forward limit. This is not always the case. We must require that $\tau$ be non-negative in the forward limit, which translates into the mass criterion
\begin{align}
\label{eq:mIneq}
m^2 \geq \frac{1}{q-1}\,.
\end{align}
The region excluded by this inequality is marked in purple in Figure \ref{fig:Disc}. For $q \geq 2-\sqrt{2}\approx 0.5858$, the inequality \eqref{eq:mIneq} is already implied by the condition $m^2\geq 1-\frac{2}{q}$ that is imposed by the pole at $s=s_1$, but for $q < 2-\sqrt{2}$, the inequality \eqref{eq:mIneq} provides a new constraint. For $\tau > 0$, the branch cut discontinuity is purely imaginary, and by inspection of equation \eqref{eq:res2} we find that the imaginary part is positive whenever
\begin{align}
0<\tau < \frac{1}{q}
\hspace{6mm}
\text{or}
\hspace{6mm}
\frac{1}{q^{2n}} < \tau < \frac{1}{q^{2n+1}}
\hspace{6mm}
\text{for }n\in\mathbb{N}\,.
\end{align}
The exclusion regions implied by these inequalities in the space of $m^2$ and $q$ are depicted in red in Figure \ref{fig:Disc}. No new regions in parameter space are excluded compared with the constraint $m^2\leq \frac{1}{3}$ that also came from the pole at $s=s_1$. 

\subsection*{$\bullet\,\,N = 1$}

\begin{figure}[h]
    \centering
\begin{align*}
\begin{matrix}\text{
		\includegraphics[width=0.4\textwidth]{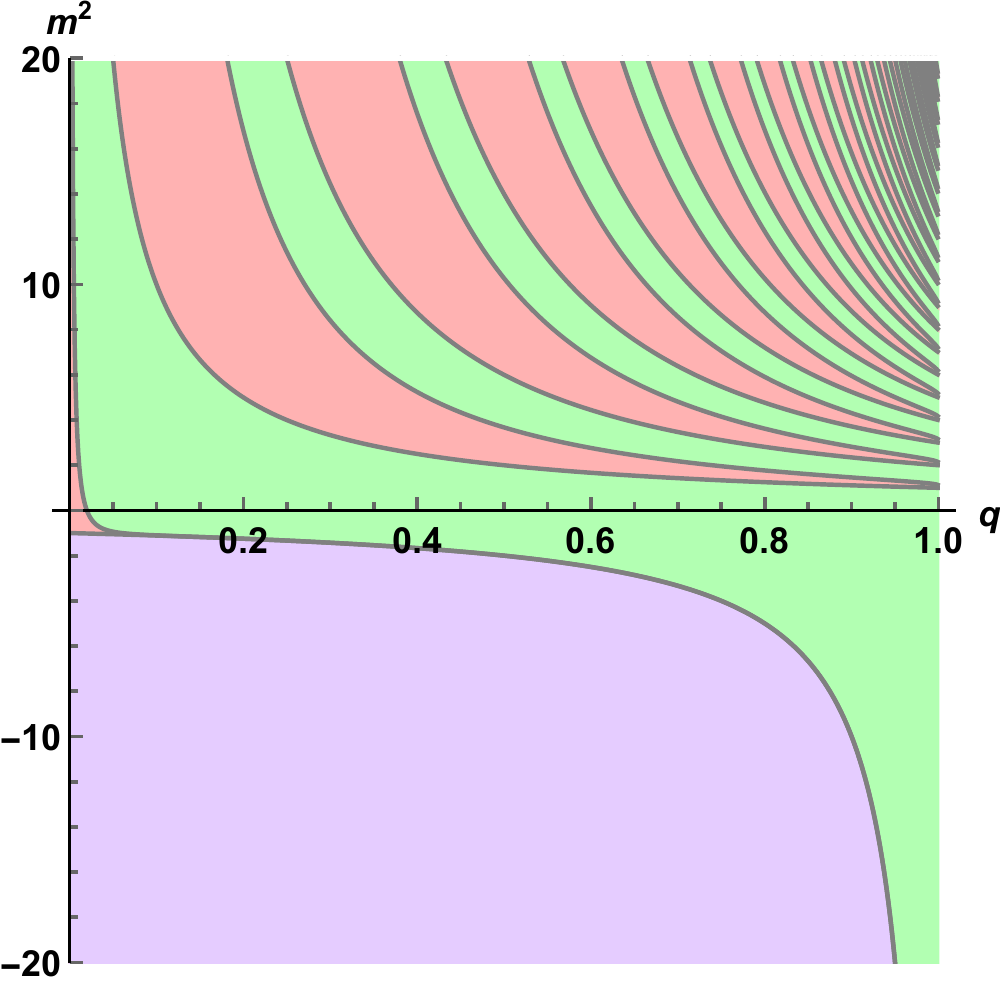}
}\end{matrix}
\hspace{14mm}
\begin{matrix}\text{
		\includegraphics[width=0.4\textwidth]{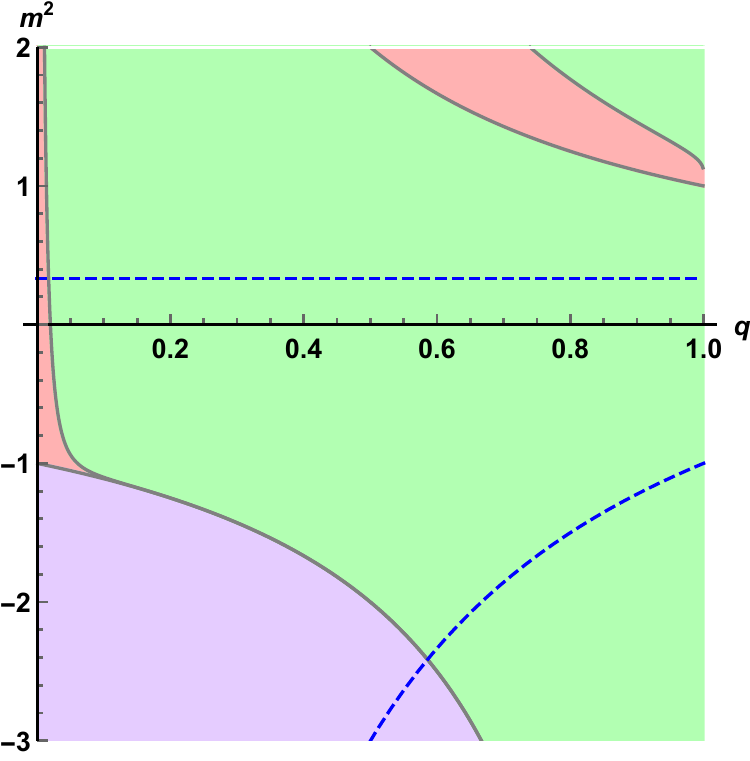}
}\end{matrix}
\end{align*}
    \caption{Green: Regions of parameter space where the derivative $\frac{d}{dt}\log\text{Disc}\big[\text{Im}\,A_q(s,t)\big]$ is positive everywhere on the branch cut. Red: Regions for which positivity is violated. Purple: Region where the branch cut is complex. The blue dotted lines in the zoomed-in plot on the right indicate the bounds $1-\frac{2}{q}\leq m^2$ and $m^2<\frac{1}{3}$ coming from the residues.}
    \label{fig:dDisc}
\end{figure}

The $t$-derivative of the branch cut discontinuity is given by
\begin{align}
&\hspace{54mm}\frac{d}{dt}\text{Disc}\Big[\text{Im}\,A_q(s,t)\Big]
=
\\ \nonumber
&\frac{q-1}{\tau}\text{Disc}\Big[\text{Im}\,A_q(s,t)\Big]
\bigg(
\frac{\log|\sigma|}{\log q}
+\cot\Big(\pi\frac{\log\tau}{\log q}\Big) 
\frac{\pi}{\log q}
+\sum_{n=0}^\infty\frac{q^n}{\tau\sigma-q^n}
-\sum_{n=0}^\infty\frac{q^n}{\tau-q^n}
\bigg)\,.
\end{align}
It is convenient to eliminate $\sigma$ and $\tau$ in favour of two new variables, call them $S$ and $T$, related to the former thus
\begin{align}
\label{eq:SandT}
S \equiv \frac{\log |\sigma|}{\log q}\,,
\hspace{20mm}
T \equiv \frac{\log \tau}{\log q}\,.
\end{align}
Additionally, let us introduce the following two functions:
\begin{align}
\label{eq:Rplus}
R_+(x) \equiv \,& \sum_{n=0}^\infty \frac{1}{1+q^{x-n}}=-\frac{\log(1-q)+\psi_q(-x+\frac{\pi i}{\log q})}{\log q}\,,
\\
\label{eq:Rminus}
R_-(x) \equiv \,& \sum_{n=0}^\infty \frac{1}{1-q^{x-n}}=-\frac{\log(1-q)+\psi_q(-x)}{\log q}\,.
\end{align}
We can now rewrite the derivative of the discontinuity as
\begin{align}
\nonumber
\frac{d}{dt}\text{Disc}\Big[\text{Im}\,A_q(s,t)\Big]
=
\frac{q-1}{\tau}\text{Disc}\Big[\text{Im}\,A_q(s,t)\Big]
\bigg(
S
+\cot(\pi T) \frac{\pi}{\log q}
-R_+(S+T)
+R_-(T)
\bigg)\,.
\end{align}
In appendix \ref{sec:App} the following two identities:
\begin{align}
\label{eq:cruc1}
R_+(x)=\,&R_+(-x-1)+x+\frac{1}{2}+ \sum_{n=1}^\infty B_n^{(+)}\sin(2\pi n x)\,,
\\
\label{eq:cruc2}
R_-(x)=\,&R_-(-x-1)+x+\frac{1}{2}
-\frac{\pi}{\log q}\cot(\pi x)
+\sum_{n=1}^\infty 
B_n^{(-)}\sin(2\pi n x)\,,
\end{align}
where the coefficients $B_n^{(\pm)}$ are given by
\begin{align}
\label{eq:Bpm}
B_n^{(+)}=
-\frac{4\pi}{\log q} \frac{e^{\frac{2\pi^2}{\log q}n}}{1-e^{\frac{4\pi^2}{\log q}n}}\,,
\hspace{20mm}
B_n^{(-)}=-\frac{4\pi}{\log q}
\frac{e^{\frac{4\pi^2}{\log q}n}}{1-e^{\frac{4\pi^2}{\log q}n}}\,.
\end{align}
Through the use of these identities, we find that
\begin{align}
\label{dDiscFinal}
&\hspace{30mm}
\bigg(\frac{q-1}{\tau}\text{Disc}\Big[\text{Im}\,A_q(s,t)\Big]\bigg)^{-1}
\frac{d}{dt}\text{Disc}\Big[\text{Im}\,A_q(s,t)\Big]
 =
\\ \nonumber
& \hspace{-5mm}
R_-(-T-1)
-R_+(-S-T-1)
+\sum_{n=1}^\infty
B_n^{(-)}
\sin\big(2\pi nT\big)
-\sum_{n=1}^\infty
B_n^{(+)}
\sin\big(2\pi n(S+T)\big)\,.
\end{align}
In the forward limit, the right-hand-side must be non-positive for all $S$. The function $R_-(x)$ is negative for all negative arguments, whereas $R_+(x)$ is positive for all (real) $x$. This means that for $m<\frac{1}{q}$, any unitarity violations will be entirely due to the coefficients $B_n^{(\pm)}$. It turns out that unless $q$ is close to zero, the coefficients $B_n^{(\pm)}$ are unnaturally small compared with naive expectations for dimensionless numbers that depend only on a single order-one number $q$. For this reason, any unitarity violations will be well-concealed. For example, when $q=\frac{1}{2}$, the leading sine coefficients are given by
\begin{align}
B^{(+)}_1\big|_{q=\frac{1}{2}} \approx 7.77 \cdot 10^{-12}\,,
\hspace{15mm}
B^{(-)}_1\big|_{q=\frac{1}{2}}\approx 3.33  \cdot 10^{-24}\,.
\end{align}
To flesh out possible positivity violations and obtain the sharpest possible inequality, we take the limit $S\rightarrow \infty$ of equation \eqref{dDiscFinal}, in the which limit the term $-R_+(-S-T-1)$ tends to zero. This limit corresponds to zooming in on the part of the branch cut that is close to the accumulation point. Additionally, we pick the non-integer part of $S$ to be such that the sum $\sum_{n=1}^\infty B_n^{(+)}\sin\big(2\pi n(S+T)\big)$ is as small as possible. We then arrive at the inequality
\begin{align}
&
R_-(-T_0-1)
+\sum_{n=1}^\infty
B_n^{(-)}
\sin\big(2\pi nT_0\big)
\leq \underset{x}{\text{min}} 
\sum_{n=1}^\infty
B_n^{(+)}
\sin\big(2\pi n x\big)\,,
\label{finalIneq}
\end{align}
where $T_0$ is the forward-limit value of $T$:
\begin{align}
T_0= \frac{\log \tau}{\log q}\Big|_{t=0}=\frac{\log \big(1+m^2(1-q)\big)}{\log q}\,.
\end{align}
For any given value of $q$, we can check the inequality \eqref{finalIneq} numerically to determine the allowed values for $T_0$ and thereby the allowed values for $m^2$. By performing this check for a large set of $q$-values between zero and one, we arrive the plot in Figure \ref{fig:dDisc}. An important aspect to note is the left-most excluded region in red. When combined with the inequality $m^2\leq\frac{1}{3}$ imposed by the pole at $s=s_1$, this exclusion region implies that the Coon amplitude violates positivity for any $m^2$ when $q<0.0172$. 

\subsection*{$\bullet\,\, N\leq 8$}
\begin{figure}[h]
    \centering
\begin{align*}
\begin{matrix}\text{
		\includegraphics[scale=1]{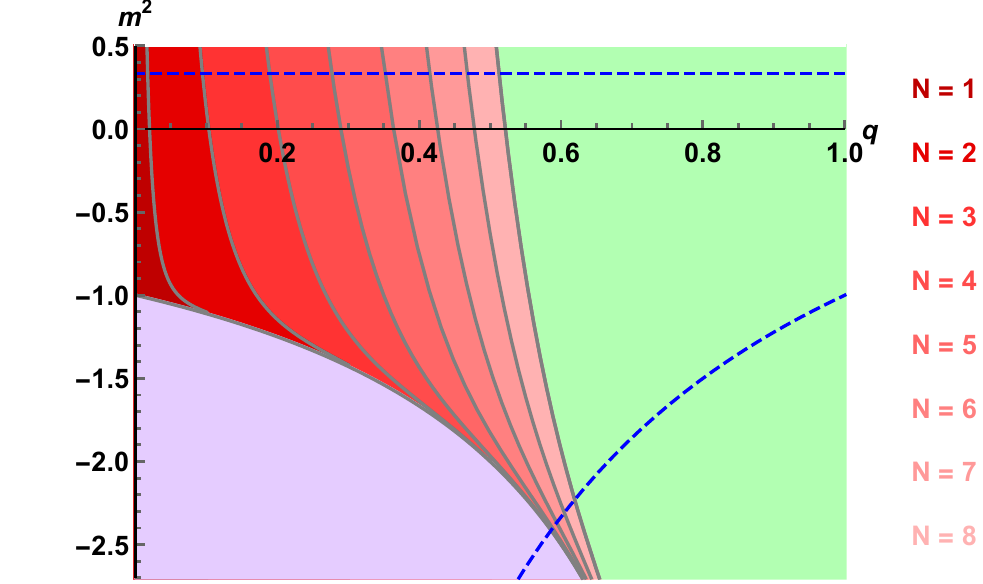}
}\end{matrix}
\end{align*}
    \caption{Red: Positivity-violating regions of parameter space determined from $N$-fold derivatives of the branch cut discontinuity for $N\leq 8$. Purple: Region where the branch cut is complex. The blue dotted lines indicate the bounds $1-\frac{2}{q}\leq m^2$ and $m^2<\frac{1}{3}$ coming from the residues.}
    \label{fig:d7disc}
\end{figure}

If we use the symbol $\overset{\divideontimes}{=}$ to indicate an identity that becomes correct in the $S\rightarrow \infty$ limit, or, more precisely, to indicate that the ratio of the left- and right-hand sides tend to one in this limit, then we found in the previous section that
\begin{align}
\label{eqWithNewSymbol}
\frac{d}{dt}\text{Disc}\Big[\text{Im}\,A_q(s,t)\Big]
 \,\overset{\divideontimes}{=}\, \frac{q-1}{\tau}\,g_1(S,T)\,\text{Disc}\Big[\text{Im}\,A_q(s,t)\Big] \,,
 \end{align}
where I have introduced the shorthand
 \begin{align}
g_1(S,T)
\equiv
R_-(-T-1)
+\sum_{n=1}^\infty
\bigg(B_n^{(-)}
\sin\big(2\pi n T\big)
-B_n^{(+)}
\sin\big(2\pi n(S+T)\big)\bigg)\,.
\end{align}
It is a simple exercise to check that
\begin{align}
&\frac{d^2}{dt^2}\text{Disc}\Big[\text{Im}\,A_q(s,t)\Big]  \,\overset{\divideontimes}{=}\,
\frac{q-1}{\tau}\,g_2(S,T)\,\frac{d}{dt}\text{Disc}\Big[\text{Im}\,A_q(s,t)\Big]\,,
\end{align}
where I have introduced another shorthand,
\begin{align}
g_2(S,T)=-1+g_1(S,T)+\frac{1}{\log q}\frac{\frac{d}{dT}g_1(S,T)}{g_1(S,T)}\,.
\end{align}
More generally, the following identity holds:
\begin{align}
\frac{d^{N+1}}{dt^{N+1}}\text{Disc}\Big[\text{Im}\,A_q(s,t)\Big]
 \,\overset{\divideontimes}{=}\,
\frac{q-1}{\tau}\,g_{N+1}(S,T)\,\frac{d^N}{dt^N}\text{Disc}\Big[\text{Im}\,A_q(s,t)\Big] \,,
\end{align}
where the function $g_N(S,T)$ satisfies the recursive relation
\begin{align}
g_{N+1}(S,T)=\,&-1+g_N(S,T)+\frac{1}{\log q}\frac{\frac{d}{dT}g_N(S,T)}{g_N(S,T)}\,.
\end{align}
Unitarity dictates that each of these functions $g_N(S,T)$ be negative for all $S$ when $T=T_0$. Since $g_N(S,T)$ is periodic in $S$ with unit periodicty, it suffices, for a given $T$, to check $g_N(S,T)$ for positivity along a unit interval. By sweeping through different values of $T$, we can generate exclusion plots like Figure \ref{fig:dDisc} for different values of $N$, although the numerics get increasingly cumbersome as $N$ increases. Figure \ref{fig:d7disc} shows the positivity-violating regions of the Coon amplitude that are excluded for $N\leq 8$. We see that with increasing values of $N$, the exclusion region grows until it covers most of parameter space. It is very conceivable that everything is excluded in the limit as $N$ goes to infinity.

\subsection{Numerical evaluation of Gegenbauer coefficients}
\label{subsec:Numeric}

\begin{figure}[h]
    \centering
\begin{align*}
\begin{matrix}\text{
		\includegraphics[scale=0.75]{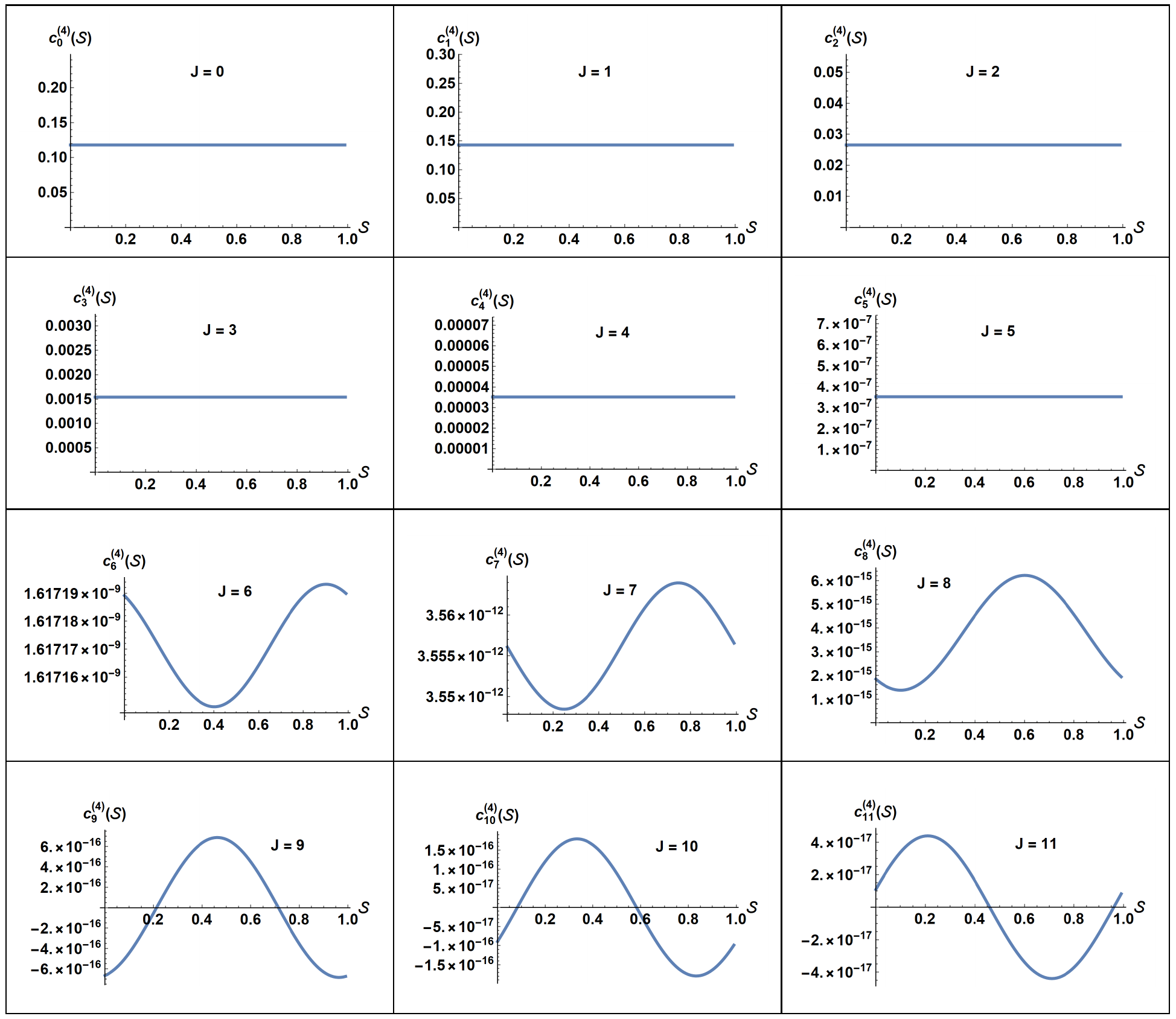}
}\end{matrix}
\end{align*}
    \caption{Asymptotic partial wave coefficients on the branch cut in the limit as $s$ approaches $s_\infty$ from above, for $d=4$, $m^2=0$, and $q=1/2$.}
    \label{fig:Geg}
\end{figure}

By integrating equation \eqref{dDiscFinal}, we can obtain an alternative functional expression for the branch cut discontinuity up to a non-zero $t$-independent function $G(s)$, whose exact form we will not need. We find that
\begin{align}
\label{eq:altDisc}
\text{Disc}\Big[\text{Im}\,A_q(s,t)\Big]
=
G(s)
F(S,T)
\prod_{n=0}^\infty
\frac{1-q^{n+T+1}}{1+q^{n+S+T+1}}\,,
\end{align}
where the function $F(S,T)$ is given by
\begin{align}
\label{eq:F(S,T)}
F(S,T)=\exp\bigg(
\sum_{n=1}^\infty \frac{\log q}{2\pi n}\Big(
B_n^{(+)}
\cos\big(2\pi n(S+T)\big)
-
B_n^{(-)}
\cos\big(2\pi nT\big)
\Big)
\bigg)\,.
\end{align}
From its definition, it is clear that $F(S,T)$ is insensitive to the integer part $\text{int}(S)$ of $S$. In other words, $F(S,T)$ depends only on $S$ through the quantity $\mathcal{S}\equiv S - \text{int}(S)$. Consider now the limit $S \rightarrow 0$, ie. the limit when $s$ tends to $s_\infty$ from above. In this limit $G(s)$ must tend to zero, since the branch cut discontinuity tends to zero, but as long as $s$ is greater than $s_\infty$ by any amount, however small, $G(s)$ will generically be non-zero, and we have the asymptotic identity
\begin{align}
\text{Disc}\Big[\text{Im}\,A_q(s,t)\Big]
 \,\overset{\divideontimes}{=}\,
G(s)
F(\mathcal{S},T)
\prod_{n=0}^\infty
(1-q^{n+T+1})\,,
\end{align}
from which we obtain an asymptotic partial wave expansion,
\begin{align}
\text{Disc}\Big[\text{Im}\,A_q\Big(s,\frac{(s_\infty-4m^2)(\cos\theta-1)}{2}\Big)\Big] \,\overset{\divideontimes}{=}\, G(s)\sum_{J=0}^\infty c_J^{(d)}(\mathcal{S})P_J(\cos\theta)\,.
\end{align}
Using the orthogonality relation for the Gegenbauer polynomials,
\begin{align}
&\int_{-1}^1 dx\,(1-x^2)^{\frac{d-4}{2}} P_J(x)P_{J'}(x) = \delta_{J,J'}\,\mathcal{N}(d,J)\,,
\\[4pt]
&\text{where }\mathcal{N}(d,J)=\frac{2^{5-d}\pi\Gamma(d+J-3)}{(d+2J-3)\,J!\,\Gamma(\frac{d-3}{2})^2}\,,
\end{align}
we can extract the asymptotic partial wave coefficients through numeric integration:
\begin{align}
c_J^{(d)}(\mathcal{S})
=\frac{1}{\mathcal{N}(d,J)}\int_{-1}^1 dx\,
(1-x^2)^{\frac{d-4}{2}}\,P_J(x)\,
F\big(\mathcal{S},T(x)\big)\prod_{n=0}^\infty
(1-q^{n+T(x)+1})\,,
 \end{align}
where $T$, defined in \eqref{eq:SandT}, depends on $x=\cos\theta$ through $\tau$, defined in \eqref{eq:sigmaAndTau}, which depends on $t$. Specifically,
\begin{align}
T(x)=\frac{1}{\log q}\log\bigg(1+m^2(1-q)+\frac{(s_\infty-4m^2)(1-x)(1-q)}{2}\bigg)\,.
\end{align}
The asymptotic coefficients $c_J^{(d)}(\mathcal{S})$ are related to the coefficients $a_J^{(d)}(s)$ in equation \eqref{eq:discDecomp} by
\begin{align}
a_J^{(d)}(s)\,\overset{\divideontimes}{=}\, G(s)\,c_J^{(d)}(\mathcal{S})\,.
\end{align}
The positivity condition \eqref{eq:coeffIneq2} on the coefficients $a_J^{(d)}(s)$ therefore implies a condition on the asymptotic coefficients $c_J^{(d)}(\mathcal{S})$: for any value of $\mathcal{S}$, the coefficients $c_J^{(d)}(\mathcal{S})$ must have the same sign for all $J$. Figure \ref{fig:Geg} shows plots of the first twelve asymptotic coefficients for the Coon amplitude with $m^2=0$ and $q=1/2$ when $d=4$. We see that the coefficients start out positive, but from $J=9$ and onward the coefficients assume both positive and negative values. By repeated numerical experiments, one finds that this behaviour is general across different values of $m^2$, $q$, and $d$: the coefficients $c_J^{(d)}(\mathcal{S})$ start out positive and of order one, but with increasing $J$ they decay and eventually reach a level where the oscillations due to the cosine terms in equation \eqref{eq:F(S,T)} induce negative values. Let us use the symbol $J_c(d,m^2,q)$ to indicate, for fixed $m^2$, $d$, and $q$, the lowest value of $J$ at which the coefficients $c_J^{(d)}(\mathcal{S})$ assume negative values. We have seen that $J_c(4,0,1/2)=9$. In Table \ref{table:1}, we list the values of $J_c$ for a variety of different masses and $q$-values in dimensions $d=4$ to $d=26$. As $q$ approaches 1, the values of $J_c$ increase as the cosine coefficients $B_n^{(+)}$ and $B_n^{(-)}$ are increasingly suppressed, but we observe that positivity is violated in every case.

\begin{table}[h!]
\centering
\scalebox{0.8}{
\renewcommand{\arraystretch}{1.5}
\begin{tabular}{|c|c|c|c|c|c|c|c|c|c|c|} 
\cline{2-11}
 \multicolumn{1}{c|}{} & $q$ & 0.1 & 0.2 & 0.3 & 0.4 & 0.5 & 0.6 & 0.7 & 0.8 & 0.9
\\
\hline
\cellcolor{lightgray}$m^2=-1$ & \multirow{3}{*}{$J_c$}  &\cellcolor{lightgray} 3 &\cellcolor{lightgray} 4 &\cellcolor{lightgray} 5 &\cellcolor{lightgray} 7 &\cellcolor{lightgray} 9 &\cellcolor{lightgray} 12 &\cellcolor{lightgray} 17 &\cellcolor{lightgray} 27 &\cellcolor{lightgray} 57
 \\
 \cline{1-1}\cline{3-11}
 \cellcolor{lightgray2}$m^2=0$\textcolor{lightgray2}{...}  &  &\cellcolor{lightgray2} 3 &\cellcolor{lightgray2} 4 &\cellcolor{lightgray2} 5
&\cellcolor{lightgray2} 6 &\cellcolor{lightgray2} 9 &\cellcolor{lightgray2} 12 &\cellcolor{lightgray2} 17 &\cellcolor{lightgray2} 27 &\cellcolor{lightgray2} 56
 \\
 \cline{1-1}\cline{3-11}
 \cellcolor{lightgray}$m^2=1/3$ &  
&\cellcolor{lightgray} 3 &\cellcolor{lightgray} 4 &\cellcolor{lightgray} 5 &\cellcolor{lightgray} 6 &\cellcolor{lightgray} 8 &\cellcolor{lightgray} 12 &\cellcolor{lightgray} 17 &\cellcolor{lightgray} 26 &\cellcolor{lightgray} 56
 \\
 \hline
\end{tabular}}
\caption{Lowest value $J_c$ of exchanged spin $J$ for which the asymptotic Gegenbauer coefficients become negative. The values of $J_c$ are the same in all dimensions from $d=4$ to $d=26$ within the ranges of parameters listed in the table.}
\label{table:1}
\end{table}

\section{Curing the Coon Amplitude?}
\label{sec:Cure}

Studies of the poles of the Coon amplitude persuasively suggest that $A_q(s,t)$ constitutes a unitary one-parameter family of amplitudes that smoothly interpolate between the Veneziano amplitude in the limit $q\rightarrow 1$ and a scalar field theory amplitude in the limit $q\rightarrow 0$. In this paper we have seen that an analysis of the branch cut of $A_q(s,t)$ presents an obstacle to this picture. For most values of $q$ between zero and one, and possibly for all, the imaginary discontinuity of the amplitude does not decompose into a positively weighted sum of partial waves everywhere along the cut, as would have been expected from unitarity. 

From the work of Refs.~\cite{klebanov2006dynamics}, \cite{alday2007gluon} and \cite{maldacena2022accumulation}, we know that an accumulation point spectrum and a scaling behaviour 
\begin{align}
\log\mathcal{A}(s,t) \sim -\,\#\log(s)\log(t)+...
\end{align}
for an amplitude $\mathcal{A}(s,t)$ at large $s$ and $t$ values---both of which are properties the Coon amplitude---are indeed features of legitimate physical 4-point scattering amplitudes. For this reason, an appealing interpretation of the positivity violations of the Coon amplitude is the following:
\begin{itemize}

   \item \emph{The Coon amplitude can be further modified into a physical and unitary amplitude.}
    
    The issue of non-polynomial residues in the original Coon amplitude was fixed through multiplication by an extra factor of $e^{\frac{\log\sigma\,\log\tau}{\log q}}$, 
    and the analytic bootstrap derivation of the Coon amplitude in Ref.~\cite{cheung2023stringy} required the introduction by hand of a suitable choice of corrective $s$- and $t$-dependent factors and expressly left open the possibility of other choices. Perhaps an extra modification of the Coon amplitude could fix the issue of negativity on the branch cut. Ideally, such a modification would not alter the residues by anything beyond constant factors. And in fact when choosing how to define the Coon amplitude, there are degrees of freedom which can adjust the branch cut without changing the signs of the residues. Consider for example the following modification:
    \begin{align*}
    \hspace{-8mm}
A_q(s,t)\rightarrow \exp\Bigg[\sum_{n=1}^\infty 
B_n\bigg(\cos\Big(2\pi n\frac{\log (\tau\sigma)}{\log q}\Big) 
-\cos\Big(2\pi n\frac{\log \sigma}{\log q}\Big) 
-\cos\Big(2\pi n\frac{\log \tau}{\log q}\Big) 
\bigg)
\Bigg]A_q(s,t)\,.
\end{align*}
    The extra exponential factor equals a positive constant at the poles of the Coon amplitude for any choice of coefficients $B_n$, and differentiating the factor pulls down trigonometric functions that could ideally be tuned to cancel the sine terms in \eqref{dDiscFinal} that were responsible for the positivity violations. However, this strategy is complicated by the fact that such a factor comes with its own imaginary discontinuity. One could avoid this complication by, instead of cosine functions in the exponent, using elliptic functions that are also periodic in the imaginary direction---except that any non-constant elliptic function must have a pole. Attempting to formulate a version of the Coon amplitude expressed in terms of quasiperiodic functions instead, it may happen that one arrives at the following as a suitable candidate amplitude:
\begin{align}
\label{eq:CoonAlt}
\hspace{-8mm}
A_q(s,t)=\frac{q-1}{\log q}\frac{\Gamma_q(1+\frac{\log(\sigma\tau)}{\log q})}{\Gamma_q(1+\frac{\log\sigma}{\log q})\Gamma_q(1+\frac{\log\tau}{\log q})}
\frac{\sigma(\frac{\log(\sigma\tau)}{\log q}|\frac{1}{2},-\frac{\pi i}{\log q})}{\sigma(\frac{\log \sigma}{\log q}|\frac{1}{2},-\frac{\pi i}{\log q})\sigma(\frac{\log \tau}{\log q}|\frac{1}{2},-\frac{\pi i}{\log q})}
e^{-2\zeta(\frac{1}{2}|\frac{1}{2},-\frac{\pi i}{\log q})\frac{\log \sigma\,\log\tau}{\log^2 q}}\,,
\end{align}
where $\zeta(z|\omega_1,\omega_2)$ and $\sigma(z|\omega_1,\omega_2)$ are Weierstrass functions associated to a lattice with periods $2\omega_1$ and $2\omega_2$. But it happens to be the case that \eqref{eq:CoonAlt} is mathematically identical to \eqref{eq:Acoon} and \eqref{eq:AcoonAgain}, as can be shown using the identity \eqref{eq:qEuler} derived in the appendix.
\end{itemize}
Some alternative interpretations of the positivity violations are listed below along with brief comments and speculations.

\begin{itemize}
    \item \emph{The Coon amplitude is physical despite the negativity its branch cut exhibits in the forward limit.} 
    
    The concept of string tree-amplitudes with accumulation point spectra is not well understood. It remains a conceivable possibility that whatever potential physical mechanism is responsible for the Coon amplitude subtly reshapes aspects of standard unitarity criteria or that negativity issues of partial amplitudes disappear at the level of full amplitudes.
    
    \item \emph{It is the original Coon amplitude $\mathcal{A}_q(s,t)$ in \eqref{Afirst}, which has no branch cut, rather than the modified Coon amplitude $A_q(s,t)$ in \eqref{eq:Acoon}, that is physical.}
    
    Certainly between non-unitarity and non-polynomial residues, the latter is the lesser of two evils. Indeed, Ref.~\cite{huang2022uv} recently argued in detail for the permissibility of non-polynomial residues in unitary amplitudes. It may also be worth noting in this context, that Ref.~\cite{fernandez2022cornering} in their exclusion plot for the Lovelace-Shapiro amplitude includes the $q$-deformed version of the Lovelace-Shapiro amplitude with non-polynomial residues and no branch cut rather than the version with a branch cut and polynomial residues. But, as Ref.~\cite{cheung2023veneziano} points out, infinite-spin exchange on an isolated pole implies the exchange of an infinitely extended object, which contravenes locality.
    
    \item \emph{The Coon amplitude is unphysical.} 
    
    One could imagine, for example, that a theory describing the amplitude requires $q$-deforming spacetime itself. Ref.~\cite{bhardwaj2022unitarity} entertains in a footnote the idea of decomposing the poles of the Coon amplitude into $q$-Gegenbauer polynomials instead of the usual Gegenbauer polynomials. From a physical perspective, this is perhaps the least interesting possibility. If such an interpretation is correct, one may ask why the partial wave analysis of the poles of $A_q(s,t)$ is indicative of unitarity.
\end{itemize}
The task to investigate and uncover what manner of interpretation comes closer to the truth is left for the future.

\subsection*{Acknowledgements}
I am grateful to Zohar Komargodski and Nikita Nekrasov for illuminating discussions and to Zohar Komargodski for incisive comments on this work.

\appendix

\section{The $q$-gamma and $q$-polygamma functions and the Weierstrass elliptic and quasiperiodic functions}
\label{sec:App}

In this appendix we establish a set of identities relating the Weierstrass functions on rectangular lattices to the $q$-gamma and $q$-polygamma functions for real $q$ between zero and one. These identities will allow us to derive equations \eqref{eq:cruc1}, \eqref{eq:cruc2} and \eqref{eq:Bpm}, which we used in the analysis of the Coon amplitude branch cut in section \ref{sec:Cut}, and also to derive a $q$-generalization of Euler's reflection formula \eqref{eq:reflection}, from which the equivalence between the Coon amplitude in \eqref{eq:Acoon} and \eqref{eq:AcoonAgain} its alternative expression given in equation \eqref{eq:CoonAlt} in section \ref{sec:Cure} follows.

In equations \eqref{eq:Rplus} and \eqref{eq:Rminus} we introduced two sum functions related to $\psi_q(z)$. From their definitions it immediately follows that
\begin{align}
\label{eq:Rrecursion}
R_\pm(x+1)=\,&R_\pm(x)+\frac{1}{1\pm q^{x+1}}\,.
\end{align}
Suppose now that $y=x + M$ with $M\in \mathbb{N}$. In that case, by applying the identity \eqref{eq:Rrecursion} $M$ times we find that
\begin{align}
R_\pm(y)=\,&R_\pm(x)+\sum_{n=1}^M \frac{1}{1\pm q^{x+n}}
\\
=\,&R_\pm(x)+\sum_{n=1}^M\Big(1 \mp \frac{q^{x+n}}{1\pm q^{x+n}}\Big)
\\
=\,&R_\pm(x)+M - \sum_{n=1}^M\frac{1}{1\pm q^{-x-n}}
\\
=\,&R_\pm(x)+M- \sum_{n=0}^\infty \frac{1}{1\pm q^{-x-1-n}}+\sum_{n=0}^\infty \frac{1}{1\pm q^{-x-M-1-n}}
\\
=\,&R_\pm(x)+y-x-R_+(-1-x) +R_\pm(-1-y)\,.
\end{align}
Rearranging terms, we see that
\begin{align}
R_\pm(y)-R_\pm(-1-y)-y = R_\pm(x)-R_\pm(-1-x)-x \,.
\end{align}
In other words, if we introduce functions $F_\pm (x)$ defined by
\begin{align}
F_+(x) \equiv\,& R_\pm(x)-R_\pm(-1-x)-x-\frac{1}{2}=\frac{\psi_q\big(x+1+\frac{\pi i}{\log q}\big)-\psi_q\big(-x+\frac{\pi i}{\log q}\big)}{\log q}-x-\frac{1}{2}\,,
\nonumber
\\
F_-(x) \equiv\,& R_\pm(x)-R_\pm(-1-x)-x-\frac{1}{2}=\frac{\psi_q(x+1)-\psi_q(-x)}{\log q}-x-\frac{1}{2}\,,
\label{eqDef}
\end{align}
then these functions are periodic with unit periodicity: 
\begin{align}
F_\pm(x+M) = F_\pm(x)\,.
\end{align}
In the definitions \eqref{eqDef}, I chose to introduce a term $-\frac{1}{2}$ because in that case $F_\pm(x)$ equals
\begin{align}
F_\pm (x)=\,&
\sum_{n=0}^\infty \bigg(\frac{1}{1\pm q^{x-n}}-\frac{1}{1\pm q^{-x-n}}\bigg)
+\frac{1}{1\pm q^{-x}} -\frac{1}{2}-x\,,
\end{align}
and by using the identity 
\begin{align}
\frac{1}{1\pm q^{-x}}-\frac{1}{1\pm q^x}=1\,,
\end{align}
we find that
\begin{align}
F_\pm(x)=\,&
\sum_{n=0}^\infty \bigg(\frac{1}{1\pm q^{x-n}}-\frac{1}{1\pm q^{-x-n}}\bigg)
+\frac{1}{2}\bigg(\frac{1}{1\pm q^{-x}}-\frac{1}{1\pm q^{x}}\bigg)-x\,,
\end{align}
from which we see that the functions $F_\pm(x)$ are odd. Since $F_+(x)$ is an odd function with period one and no poles (for real $x$), it admits a Fourier expansion in sine functions:
\begin{align}
F_+(x)=\sum_{n=1}^\infty B_n^{(+)}\sin(2\pi n x)\,,
\hspace{15mm}\text{where }
B_n^{(+)}=2\int_0^1dx\,\sin(2\pi n x)F_+(x)\,.
\label{eq:BplusInt}
\end{align}
Meanwhile, the function $F_-(x)$ has poles with residue $-(\log q)^{-1}$ at all integer arguments, so that its Fourier expansion can be written as
\begin{align}
\label{eq:FmFourier}
&F_-(x)=\-\frac{\pi}{\log q}\cot(\pi x)
+\sum_{n=1}^\infty 
B_n^{(-)}\sin(2\pi n x)\,,
\\
\label{eq:BminusInt}
&\text{where }
B_n^{(-)}=2\int_0^1dx\,\sin(2\pi n x)\Big(F_+(x)+\frac{\pi}{\log q}\cot(\pi x)\Big)\,.
\end{align}
It should be possible to directly evaluate the integrals in \eqref{eq:BplusInt} and \eqref{eq:BminusInt} and thereby obtain equations \eqref{eq:Bpm}. But because these integrals are not so easy to carry out, we will adopt a different approach and instead relate $F_\pm(x)$ to a function with a known Fourier expansion. To this end, we observe that it follows immediately from the definitions \eqref{eq:Rplus} and \eqref{eq:Rminus} of the functions $R_+(x)$ and $R_-(x)$ that, allowing them to assume complex arguments, they are periodic in the imaginary direction with periodicity $\frac{2\pi i}{\log q}$. Therefore, their derivatives $R'_+(x)$ and $R'_-(x)$ are similarly periodic, and likewise are the functions
\begin{align}
F'_\pm(z) = R'_\pm(z)+R'_\pm(-1-z)-1\,.
\end{align}
Furthermore, since $F_\pm(z)=F_\pm(z+1)$, the functions $F'_\pm(z)$ are also periodic in the real direction with unit periodicity. We conclude that $F'_+(z)$ and $F'_-(z)$ are elliptic functions. We also note that $F'_-(z)$ has the Laurent expansion
\begin{align}
\label{eq:F'expansion}
F_-'(z)
=\,&
\frac{1}{\log q\,z^2}
+
\frac{2\psi_q^{(1)}(1)}{\log q}
-1
-\frac{\log q}{12}
+
\bigg(
\frac{\psi_q^{(3)}(1)}{\log q}
+\frac{\log^3 q}{240}
\bigg)z^2+
\bigg(
\frac{\psi_q^{(5)}(1)}{12\log q}
-\frac{\log^5 q}{6048}
\bigg)z^4
+
...
\end{align}
Consider now the Weierstrass elliptic function $\wp(z|\omega_1,\omega_2)$ with complex half-periods $\omega_1$ and $\omega_2$. This function has the Laurent expansion
\begin{align}
\label{eq:wpExpansion}
\wp(z|\omega_1,\omega_2)
=\,& \frac{1}{z^2}
 +\frac{g_2(\omega_1,\omega_2)}{20}z^2
 +\frac{g_3(\omega_1,\omega_2)}{28}z^4
 +...\,,
\end{align}
where the invariants are given by
\begin{align}
g_2(\omega_1,\omega_2)
=60\hspace{-4mm}\sum_{\substack{m,n\in \mathbb{Z}\\ \{m,n\}\neq \{0,0\}}}
\frac{1}{(2m\omega_1+2n\omega_2)^4}\,,
\hspace{10mm}
g_3(\omega_1,\omega_2)
=140\hspace{-4mm}\sum_{\substack{m,n\in \mathbb{Z}\\ \{m,n\}\neq \{0,0\}}}
\frac{1}{(2m\omega_1+2n\omega_2)^6}\,.
\end{align}
By picking half-periods $\omega_1=\frac{1}{2}$ and $\omega_2=-\frac{\pi i}{\log q}$, $\wp(z|\omega_1,\omega_2)$ has the same periodicites as $F_-'(z)$. Comparing the Laurent expansions \eqref{eq:F'expansion} and \eqref{eq:wpExpansion}, we see that the function
\begin{align}
\label{eq:const}
\log q\, F'_-(z)-\wp\Big(z\Big|\frac{1}{2},-\frac{\pi i}{\log q}\Big)
\end{align}
is an elliptic function with no poles and therefore, by Liouville's theorem, \eqref{eq:const} must be a constant. The value of the constant is simply the constant term in \eqref{eq:F'expansion}. We conclude that 
\begin{align}
\label{eq:F'formula}
F'_-(z)=
\frac{1}{\log q}\wp\Big(z\Big|\frac{1}{2},-\frac{\pi i}{\log q}\Big)
+
\frac{2\psi_q^{(1)}(1)}{\log q}
-1
-\frac{\log q}{12}\,,
\end{align}
and we additionally infer that
\begin{align}
\label{eq:g23}
g_2\Big(\frac{1}{2},-\frac{\pi i}{\log q}\Big)=
20\psi_q^{(3)}(1)
+\frac{\log^4 q}{12}\,,
\hspace{10mm}
g_3\Big(\frac{1}{2},-\frac{\pi i}{\log q}\Big)=
\frac{7\psi_q^{(5)}(1)}{3}
-\frac{\log^6 q}{216}\,.
\end{align}
Recall now that the Weierstrass zeta function $\zeta(z|\omega_1,\omega_2)$ is defined by
\begin{align}
\frac{d\zeta(z|\omega_1,\omega_2)}{dz}
=-\wp(z|\omega_1,\omega_2)\,,
\hspace{15mm}
\lim_{z\rightarrow 0}
\bigg(\zeta(z|\omega_1,\omega_2)-\frac{1}{z}\bigg)=0\,.
\end{align}
With the help of this definition, we can integrate \eqref{eq:F'formula} to find that
\begin{align}
\label{eq:FmFormula}
F_-(z)=
-\frac{1}{\log q}\zeta\Big(z\Big|\frac{1}{2},-\frac{\pi i}{\log q}\Big)
+
\bigg(\frac{2\psi_q^{(1)}(1)}{\log q}
-1
-\frac{\log q}{12}
\bigg)z\,.
\end{align}
The Weierstrass zeta function has a known trigonometric expansion:\footnote{See for example the NIST Digital Library of Mathematical Functions $\S$23.8(i).}
\begin{align}
\zeta(z|\omega_1,\omega_3) = \frac{\zeta(\omega_1,\omega_1,\omega_3)z}{\omega_1}
+\frac{\pi}{2\omega_1}\cot\Big(\frac{\pi z}{2\omega_1}\Big)
+\frac{2\pi}{\omega_1}\sum_{n=1}^\infty 
\frac{(e^{i\pi\omega_3/\omega_1})^{2n}}{1-(e^{i\pi\omega_3/\omega_1})^{2n}}
\sin\Big(\frac{n\pi z}{\omega_1}\Big)\,,
\end{align}
and this expansion allows us to directly read off the Fourier coefficients $B^{(-)}_n$, while the coefficients $B^{(+)}_n$ can be obtained through the use of the identity $F_+(x)=F_-(x+\frac{\pi i}{\log q})+\frac{\pi i}{\log q}$. We thereby recover equations \eqref{eq:Bpm}, which concludes the derivation of the identities needed for section \ref{sec:Cut}.

With equation \eqref{eq:FmFormula} in hand, we can derive a $q$-generalization of Euler's reflection formula \eqref{eq:reflection} with little extra effort. Recalling the definition of the Weierstrass sigma function $\sigma(z|\omega_1,\omega_3)$,
\begin{align}
\frac{d}{dz}\log\sigma(z|\omega_1,\omega_3)
=\zeta(z|\omega_1,\omega_3)\,,
\hspace{15mm}
\lim_{z\rightarrow 0}
\frac{\sigma(z|\omega_1,\omega_3)}{z}=1\,,
\end{align}
and the relation between $F_-(z)$ and the $q$-digamma function $\psi_q(z)$ in \eqref{eqDef}, we can integrate \eqref{eq:FmFormula} and exponentiate to arrive at the formula
\begin{align}
\label{eq:qEuler}
\Gamma_q(x+1)\Gamma_q(-x)
=\,&
\frac{1-q}{\log q}
\frac{q^{x(x+1)/2}}{\sigma\Big(x\Big|\frac{1}{2},-\frac{\pi i}{\log q}\Big)
}
\exp\bigg[
\zeta\Big(\frac{1}{2}\Big|\frac{1}{2}, -\frac{\pi i}{\log q}\Big)x^2\bigg]\,.
\end{align}
See Ref.~\cite{mezHo2013q} for an equivalent formula written in terms of the Jacobi theta function.

We can also invert equations 
\eqref{eq:F'formula},
\eqref{eq:g23}, 
\eqref{eq:FmFormula}, and 
\eqref{eq:qEuler}
and use the homogeneity properties of the Weierstrass functions and invariants,
\begin{align}
g_2(\omega_1,\omega_2)
=\,&\mu^4g_2(\mu\omega_1,\mu\omega_2)\,, \label{eq:first}
\\
g_3(\omega_1,\omega_2)
=\,&\mu^6g_3(\mu\omega_1,\mu\omega_2) \,,
\\
\wp(z|\omega_1,\omega_2)
=\,& \mu^2\,\wp(\mu z\,|\,\mu \omega_1,\mu \omega_2)\,, 
\\
\zeta(z|\omega_1,\omega_2)
=\,& \mu\,\zeta(\mu z\,|\,\mu \omega_1,\mu \omega_2)\,, 
\\
\sigma(z|\omega_1,\omega_2)
=\,& \frac{1}{\mu}\,\sigma(\mu z\,|\,\mu \omega_1,\mu \omega_2)\,, \label{eq:last}
\end{align}
to arrive at the following formulas for Weierstrass functions and invariants on general rectangular lattices in terms of $q$-gamma and $q$-polygamma functions:
\begin{align}
g_2(\omega_1, |\omega_2|i)=\,&
\frac{5}{4\omega_1^4}\psi_q^{(3)}(1)
+\frac{\pi^4 }{12 |\omega_2|^4}
\,,
\\
g_3(\omega_1, |\omega_2|i)=\,&
\frac{7\psi_q^{(5)}(1)}{192\omega_1^6}
-\frac{\pi^6}{216|\omega_2|^6}
\,,
\\
\wp(z,\omega_1, |\omega_2|i)
=\,&
\frac{\psi^{(1)}_q\Big(\frac{z}{2\omega_1}+1\Big)+\psi^{(1)}_q\Big(-\frac{z}{2\omega_1}\Big)
-2\psi_q^{(1)}(1)}{4\omega_1^2}
+\frac{\pi^2}{12|\omega_2|^2}
\,,
\\
\zeta(z,\omega_1, |\omega_2|i)
=\,&
\frac{\psi_q\Big(-\frac{z}{2\omega_1}\Big)
-\psi_q\Big(\frac{z}{2\omega_1}+1\Big)}{2\omega_1}
-\frac{\pi}{2|\omega_2|}
+\Big(\frac{\psi_q^{(1)}(1)}{2\omega_1^2}
-\frac{\pi^2}{12|\omega_2|^2}
\Big)z
\,,
\\
\sigma(z, \omega_1,  |\omega_2| i)=\,&
\frac{(q-1)|\omega_2|}{\pi\,\Gamma_q\Big(\frac{z}{2\omega_1}+1\Big)\Gamma_q\Big(-\frac{z}{2\omega_1}\Big)
}
\exp\bigg[\Big(\frac{\psi_q^{(1)}(1)}{4\omega_1^2}-\frac{\pi^2}{24|\omega_2|^2}\Big)z^2-\frac{\pi}{2|\omega_2|}z\bigg]\,,
\end{align}
where $q=e^{-2\pi \omega_1/|\omega_2|}$ throughout.\footnote{In the math literature on Weierstrass functions, it is not uncommon to use the symbol $q$ to instead denote a slightly different function of the half-periods: $e^{i\pi \omega_2/\omega_1}$. I hope this difference of convention will not be a source of confusion to the reader.}

Before ending this appendix, allow me to present another set of identities that follow from the relation between the $q$-polygamma and Weierstrass functions. In the case of the standard $(q=1)$ reflection formula for the gamma function, equation \eqref{eq:reflection}, if we compare the Taylor series expansions of the left- and right-hand sides, we arrive at an equation for  $\psi^{(2n-1)}(1)$ with $n\in \mathbb{N}$. From this equation, one can derive Euler's formula for the values of the Riemann zeta function at the even positive integers:
\begin{align}
\label{eq:zetaValues}
\zeta(2n)=\frac{\psi^{(2n-1)}(1)}{(2n-1)!}=\frac{(-1)^{n+1}B_{2n}(2\pi)^{2n}}{2(2n)!}\,,
\end{align}
where $B_{2n}$ are Bernoulli numbers. In the case of the $q$-deformed identities, if we equate the Taylor expansions of the left- and right-hand sides of equation \eqref{eq:FmFourier} we find that
\begin{align}
\psi_q^{(1)}(1)
=\,&\frac{\pi^2}{6}+\frac{\log q}{2}+\frac{\log^2 q}{24}-2\pi\sum_{n=1}^\infty  
\frac{e^{\frac{4\pi^2}{\log q}n}}{1-e^{\frac{4\pi^2}{\log q}n}}2\pi n\,,
\\ \nonumber
\psi_q^{(2m+1)}(1)=\,&\frac{B_{2m+2}}{2(2m+2)}\bigg((-1)^m(2\pi)^{2m+2}+(\log q)^{2m+2}\bigg)
-(-1)^{m}2\pi\sum_{n=1}^\infty  
\frac{e^{\frac{4\pi^2}{\log q}n}}{1-e^{\frac{4\pi^2}{\log q}n}}(2\pi n)^{2m+1}
\,,
\end{align}
where $m\in \mathbb{N}$. The infinite sums on the right-hand sides are themselves equal to deformed polygamma functions, 
\begin{align}
\sum_{n=1}^\infty  
\frac{e^{\frac{4\pi^2}{\log q}n}}{1-e^{\frac{4\pi^2}{\log q}n}}(2\pi n)^{2m-1}
=
\frac{1}{2\pi}\Big(\frac{\log q}{2\pi}\Big)^{2m}\,
\psi^{(2m-1)}_{e^{\frac{4\pi^2}{\log q}}}(1)\,.
\end{align}
Unlike the $q=1$ case, then, we do not get formulas for individual $q$-polygamma functions with unit arguments, but rather formulas that relate $q$- and $q'$-deformed functions with $q'=\exp(4\pi^2/\log q)$:
\begin{align}
\label{eq:qPsi1}
&\frac{\psi_q^{(1)}(1)}{\log q}+\frac{\psi^{(1)}_{q'}(1)}{\log q'}
=\frac{\log q+\log q'+12}{24}\,,
\\[4pt] \label{eq:qPsim}
&\frac{\psi_q^{(2m+1)}(1)}{\log^{m+1} q}
+(-1)^{m}\frac{\psi^{(2m+1)}_{q'}(1)}{\log^{m+1}q'}
=\frac{B_{2m+2}}{2(2m+2)}\bigg((-1)^m\log^{m+1} q'+\log^{m+1} q\bigg)
\,,
\\[6pt]\nonumber
&\text{both provided that }\log q\,\log q'=4\pi^2\,,
\end{align}
where $m\in \mathbb{N}$. A special case occurs when $q=q'=\exp(-2\pi)$. In this case we obtain the identities
\begin{align}
\label{eq:q'Psi1}
\psi^{(1)}_{e^{-2\pi}}(1)=\,&\frac{\pi^2}{6}-\frac{\pi}{2}\,,
\\
\psi_{e^{-2\pi}}^{(4m+1)}(1)
=\,& 
\psi^{(4m+1)}(1)
=(4m+1)!\zeta(4m+2)
=\frac{B_{4m+2}(2\pi)^{4m+2}}{2(4m+2)}\,,
\end{align}
where $m\in \mathbb{N}$. From the perspective of elliptic functions, the relations \eqref{eq:qPsi1} and \eqref{eq:qPsim} owe to the fact that the rectangular lattice that governs the corresponding Weierstrass functions is invariant under $\omega_1\leftrightarrow \omega_2$. The case $q=q'=\exp(-2\pi)$ corresponds to a lemniscatic lattice, $\omega_2=i\omega_1$. It is possible to generalize \eqref{eq:first} to \eqref{eq:last} to non-rectangular lattices by allowing complex values of $q$. In this case the invariance of the lattice under modular transformations of the lattice generators must be reflected in more general relations between $q$-polygamma functions.

One way to define a $q$-deformed Riemann zeta function is as follows:\footnote{There are at least three distinct definitions of a $q$-deformed Riemann zeta function in the math literature, see Refs.~\cite{cherednik2001q}, \cite{kaneko2003variation}, and \cite{fitouhi2006mellin}. The version in equation \eqref{eq:qRiemann} is that of Ref.~\cite{kaneko2003variation}.}
\begin{align}
\label{eq:qRiemann}
\zeta_q(s)
=\sum_{n=1}^\infty\frac{q^{n(s-1)}}{[n]_q^s}
=(1-q)^s\sum_{n=1}^\infty\frac{q^{n(s-1)}}{(1-q^n)^s}\,.
\end{align}
For $q\neq 1$, the relation between the values of $\zeta_q(m+1)$ and $\psi_q^{(m)}(1)$ are not as simple as equation \eqref{eq:zetaValues}, with the exception of the identity
\begin{align}
\label{eq:zetaq2}
\zeta_q(2)=\frac{(1-q)^2}{\log^2q}\psi_q^{(1)}(1)\,.
\end{align}
For example the formula for $\zeta_q(3)$ reads
\begin{align}
\zeta_q(3)=\frac{(1-q)^3}{2\log^3q}\bigg(\psi_q^{(2)}(1)-\log q\,\psi_q^{(1)}(1)\bigg)\,.
\end{align}
By repeatedly differentiating \eqref{eq:Rminus} one can obtain formulas for $\zeta_q(n)$ for any positive integer $n$, but these formulas get increasingly lengthy. Returning to $\zeta_q(2)$, we can use equation \eqref{eq:qPsi1} in conjunction with \eqref{eq:zetaq2} to relate $\zeta_q(2)$ and $\zeta_{q'}(2)$:
\begin{align}
\frac{\log q}{(1-q)^2}\,\zeta_q(2)
+\frac{\log q'}{(1-q')^2}\,\zeta_{q'}(2)
=\frac{\log q+\log q'+12}{24}
\hspace{9mm}\text{ when }\log q\,\log q'=4\pi^2\,,
\end{align}
 and we can solve the $q$-Basel problem in the special case when $q=q'=e^{-2\pi}$:
\begin{align}
\zeta_{e^{-2\pi}}(2)=\frac{(1-e^{-2\pi})^2}{8}\Big(\frac{1}{3}-\frac{1}{\pi}\Big)\,.
\end{align}

\bibliographystyle{ssg}
\bibliography{Coon}

\end{document}